\documentclass[
aip,
jap,
amsmath,
amssymb,
reprint,
floatfix,
]{revtex4-2}

\usepackage[british]{babel}
\usepackage{csquotes}
\usepackage{graphicx}

\newcommand{\reference}[1]{\par\noindent\parbox[b]{\hsize}{
						   \everypar{\hangindent=1pc \hangafter=1}
						   #1}\par
					   }

\begin{document}	
	\title{Measurable structure factors of dense dispersions containing polydisperse, optically inhomogeneous particles}
	\author{Joel Diaz Maier}
	\affiliation{Institut für Chemie, Universität Rostock, 18051 Rostock, Germany}
	\author{Katharina Gaus}
	\affiliation{Institut für Chemie, Universität Rostock, 18051 Rostock, Germany}
	\author{Joachim Wagner}
	\email{joachim.wagner@uni-rostock.de}
	\affiliation{Institut für Chemie, Universität Rostock, 18051 Rostock, Germany}
		
	\begin{abstract}
		We exemplarily investigate how optical properties of single scatterers in interacting multi-particle systems influence measurable structure factors. Both particles with linear gradients of their scattering length density and core-shell structures evoke characteristic deviations between the weighted sum $\langle S(Q)\rangle$ of partial structure factors in a multicomponent system and experimentally accessible, measurable structure factors $S_{\mathrm{M}}(Q)$. While $\langle S(Q)\rangle$ contains only structural information of self-organising systems, $S_{\mathrm{M}}(Q)$ additionally is influenced by optical properties of their constituents resulting in features such as changing amplitudes, additional peaks in the low wavevector region or splitting of higher-order maxima which are not related to structural reasons. Hence, a careful data analysis regarding size-distribution and optical properties of single scatters is mandatory to avoid a misinterpretation of measurable structure factors. 
	\end{abstract}
	
	\maketitle
	
	\section{Introduction}
	Colloidal dispersions attract wide interest in condensed matter physics as highly tunable model systems, mimicking atoms and molecules on the much larger, mesoscopic scale with typical length scales between 10\,nm to  1000\,nm. Studying these systems enabled major advances in the comprehension of the characteristics of simple fluids and solids and in return stimulated the progress of significant theoretical developments towards the understanding of complex systems and materials (Lu \& Weitz, 2013). 
	
	Scattering experiments serve as essential methods for structural and dynamical investigations in colloidal many-particle systems (Li \emph{et al.}, 2016). Small-angle scattering (SANS with neutrons or SAXS with X-rays as a probe, respectively) enables the characterisation of colloidal suspensions across the entire range of relevant scattering vectors~$Q$ (Glatter, 2018). Employing visible light, which is also a natural choice since its wavelength is of the same order of magnitude as the typical size of a colloidal particle, the same type of analysis is in principle also possible in a simpler laboratory setup. This is however connected with the cost of a limited resolution and, as a consequence thereof, the restriction to comparatively large structures (Bohren \& Huffmann, 1983).
	
	In non-interacting systems, the positions and orientations of the colloidal particles are completely uncorrelated. Thus, the scattered intensity results solely from the superposition of the scattering functions of the single constituents. Contrary, when the particles do interact, higher-level structures emerge from inherent self-organisation due to interparticle forces, such as electrostatic and steric interactions or van der Waals attractions. The intensity is then influenced both by the optical properties of the scatterers themselves and the spatial correlations between them. For idealised, monodisperse systems, where all particles are assumed to be identical, the two contributions can be rigorously separated into the form factor $P(Q)$, containing the single-particle properties and the structure factor $S(Q)$, which encodes the structural correlations, employing the well-known factorisation $I(Q)\propto P(Q)\,S(Q)$ (Hansen \emph{et al.}, 1991). 
	
	Realistic dispersions typically exhibit a distribution of characteristics, prominently through particle size. In polydisperse, interacting systems, the characterisation via scattering experiments is in general significantly more complicated, as the factorisation of the intensity into form factor and structure factor can no longer be employed in a straightforward way (Salgi \& Rajagopalan, 1993). Additionally, the observed diffraction patterns become increasingly featureless for broader size distributions, further obstructing the interpretation of experimental intensities. The analysis of multi-component systems thus requires a thorough understanding of the underlying distributions of scattering properties and particle interactions. Insights can be gained through contrast-variation techniques (Ballauff, 2001): Selectively altering the contrast between specific particle types or between particles and the surrounding medium allows for the isolation and probing of distinct species, aiding the validation of theoretical models. 
	
	From an experimental standpoint, it is useful to analyse the so-called measurable structure factor $S_{\mathrm{M}}(Q)$, which is defined in such a way that the factorisation property $I(Q)\propto P(Q)\,S_{\mathrm{M}}(Q)$ is recovered also in the polydisperse case (Hansen \emph{et al.}, 1991). $S_{\mathrm{M}}(Q)$ is comparatively easy to access experimentally from the ratio between the intensities of an interacting suspension and a highly diluted, non-interacting one. It is as such widely used as a measure for structural correlations also in polydisperse systems, where especially the height of the principal peak is well-established as an order parameter (Banchio \emph{et al.}, 1998). Under specific circumstances, this type of analysis can however turn into a serious pitfall: $S_{\mathrm{M}}(Q)$ is fundamentally also affected by optical properties of the particles and not only by their interactions. 
	
	For certain types of dispersions, some simplifying assumptions can be employed. In dilute suspensions of strongly interacting charged particles as an example, the interparticle distances typically are about an order of magnitude larger than particle sizes because of the large electrostatic repulsion (Hayter \& Penfold, 1981). In such a case, the correlation between the particle positions and the scattering amplitudes can be neglected. This neglect of correlations leads to the so-called \enquote{decoupling approximation,} whereunder $S_{\mathrm{M}}(Q)$ can be decomposed into a structure factor that genuinely represents the averaged structural correlations and weighting factors solely dependent on the scattering amplitudes (Pusey \emph{et al.}, 1982; Kotlarchyk \& Chen, 1983). This type of analysis was also recently used in a SANS-study of moderately concentrated poly(\textit{N}-isopropylacrylamide) microgels (Zhou~\emph{et al.}, 2023), where again the importance of an accurate treatment of polydispersity was stressed.
	
	In highly concentrated suspensions, where particles are in close contact, such an approximation is no longer valid, as in these systems the correlation lengths of the particles' centre of masses are comparable to the correlation lengths inside the particles themselves. At such high particle volume fractions, excluded volume effects are the predominant contribution to the total interaction potential. The fundamental interactions in dense colloidal dispersions consisting of spherical particles can to a good approximation be theoretically described with the hard sphere model (Kirkwood \& Boggs, 1942; Widom, 1967). The radial distribution functions of an \mbox{$n$-component} mixture of hard spheres can be calculated within the Percus-Yevick closure of the Ornstein-Zernike equation (Percus \& Yevick, 1958) using Baxter's technique (Baxter, 1970) giving access to the corresponding partial structure factors (Vrij, 1978, 1979; Blum \& Stell, 1979, 1980). Building on Vrij's work (Vrij 1979; van Beurten \& Vrij, 1981), Griffith \emph{et al.} (1987) presented an analytical scattering function of a polydisperse hard-sphere fluid with a Schulz-Flory distribution (Flory, 1936; Schulz, 1939) of particle diameters. Despite very helpful, these expressions are not widely used because of their perceived complexity. Nayeri \emph{et al.} (2009) later extended this approach to core-shell structured hard spheres and used their expressions to describe experimental intensities of a hard-sphere-like microemulsion system.  Only recently, Botet \emph{et al.} (2020) provided expressions for $S_{\mathrm{M}}(Q)$ in a simple, accessible form and for a number of commonly encountered size distributions. Their analytical expressions are valid for hard, optically homogeneous spheres. 
	
	This resurge of interest is an incentive to systematically examine how different form factor models affect the characteristics of measurable structure factors. The purpose of this contribution is to raise awareness on how the particles' optical properties influence the shape of $S_{\mathrm{M}}(Q)$ while the underlying interactions remain unchanged and under which circumstances such a structure factor can still serve as a valid order parameter. We show typical examples of shapes which can be realistically encountered during contrast variation experiments, so even without explicitly employing theoretical models from this contribution or from the existing literature, a qualitative assessment of experimental findings is possible. We exemplarily analyse two simplified models for optically inhomogeneous particles: Those with a linear gradient of the scattering contrast and spheres with a core-shell structure. Nevertheless, the approach is readily adaptable to any model and provides a toolbox for the modelling of measurable structure factors for hard-sphere suspensions with arbitrary form factors. 
	
	\section{Scattering of hard-sphere mixtures}
	We consider a mixture of spherical particles, where each particle can be categorised into one of $n$ species. The composition of the mixture is specified by the number fractions $x_\alpha=N_\alpha/N$, where $N$ is the total number of particles and $N_\alpha$ is the number of particles belonging to species $\alpha$. We further restrict ourselves to elastic, single scattering events where the Born approximation is applicable. In such a case, the mean intensity
	\begin{align}\label{eqn:intensity}
		I(Q) \propto \sum\limits_{\alpha, \beta} (x_\alpha x_\beta)^{1/2} f_\alpha(Q) f_\beta(Q) S_{\alpha\beta}(Q)
	\end{align}
	is proportional to the weighted sum of the single-particle scattering amplitudes $f_\alpha(Q)$ and the partial structure factors $S_{\alpha\beta}(Q)$ (Salgi \& Rajagopalan, 1993). Herein, the scattering amplitude
	\begin{align}\label{eqn:scattering-amplitude}
		f_\alpha(Q) = 4\pi \int\limits_0^\infty \rho_\alpha(r) r^2 \frac{\sin(Q r)}{Q r} \mathrm d r
	\end{align}
	is the Fourier-Bessel transform of the scattering contrast $\rho_\alpha(r)$ whereas the partial structure factors $S_{\alpha\beta}(Q)$ are obtained from the solution of the multicomponent Ornstein-Zernike equation. Expressions for $S_{\alpha\beta}(Q)$ of the hard-sphere fluid within the Percus-Yevick closure are given by Vrij (1979), but for convenience of the reader, the solution is re-articulated in Appendix A, presented in a manner that is accessible and easily applicable. 
	
	For non-interacting particles, the partial structure factors are simply $S_{\alpha\beta}(Q) = \delta_{\alpha\beta}$, where $\delta_{\alpha\beta}$ denotes the Kronecker symbol. Eq.~(\ref{eqn:intensity}) then reduces to the size-average of the squared scattering amplitudes,
	\begin{align}\label{eqn:average-squared-scattering-amplitude}
		I(Q) \propto \langle f^2(Q) \rangle = \sum\limits_{\alpha} x_\alpha f^2_\alpha(Q).
	\end{align}
	The average form factor
	\begin{align}\label{eqn:average-form-factor}
		P(Q) = \frac{\langle f^2(Q) \rangle}{\langle f^2(0) \rangle}.
	\end{align}
	is familiarly obtained from the normalisation to forward scattering. As the measurable structure factor should satisfy the relation $I(Q)\propto P(Q)\,S_{\mathrm{M}}(Q)$, the expression
	\begin{align}\label{eqn:measurable-structure-factor}
		S_{\mathrm{M}}(Q) = \left[ \langle f^2(Q) \rangle \right]^{-1} \sum\limits_{\alpha, \beta} (x_\alpha x_\beta)^{1/2} f_\alpha(Q) f_\beta(Q) S_{\alpha\beta}(Q)
	\end{align}
	results from the combination of Eq.~(\ref{eqn:intensity}) and (\ref{eqn:average-squared-scattering-amplitude}). The averaged structure factor
	\begin{align}\label{eqn:average-structure-factor}
		\langle S(Q) \rangle = \sum\limits_{\alpha, \beta} (x_\alpha x_\beta)^{1/2} S_{\alpha\beta}(Q)
	\end{align}
	provides information about the total spatial correlations between all present particles, regardless of their species labels. It represents a true thermodynamic average independent of any optical properties. Any deviation between $S_{\mathrm{M}}(Q)$ and $\langle S(Q) \rangle$ is thus a measure for the perturbation of $\langle S(Q) \rangle$ caused by the scattering amplitudes.
	
	We now want to explore the influence of the scattering amplitudes on the shape of $S_{\mathrm{M}}(Q)$. The aim is to gain a qualitative understanding of generic patterns, so to keep the analysis tractable, only a single representative size distribution is considered. For this purpose, the Schulz-Flory distribution with probability density 
	\begin{align}\label{eqn:schulz-flory-distribution}
		c(R) = \frac{1}{\Gamma(Z+1)} \left(  \frac{Z+1}{\langle R \rangle} \right)^{Z+1} R^Z \exp\left( -\frac{Z+1}{\langle R \rangle} R \right)
	\end{align}
	is chosen. Here, $R$ is the particle radius with mean $\langle R \rangle$ and $\Gamma(x)$ represents the Gamma function. The polydispersity $p$ of the system is specified by the shape parameter $Z$ via $p^2 = (\langle R^2 \rangle - \langle R \rangle^2) / \langle R \rangle^2 =1/(Z+1)$. The idea is now to discretise the distribution to a representative $n$-component mixture. For the Schulz-Flory distribution, an efficient way to achieve this is by exploiting the generalised Gauss-Laguerre quadrature rule specifically used to calculate integrals with a weighting function like Eq.~(\ref{eqn:schulz-flory-distribution}). (D'Aguanno, 1992, 1993; Olver \emph{et al.}, 2010) The nodes and weights generated by such a procedure are equivalent to the particle radii and number fractions of a discrete mixture which shares the first $2n-1$ moments $\langle R^n \rangle$ with the original continuous distribution. For each calculated scattering function, we carefully checked that the number of nodes necessary for convergence was reached. The numerical scheme was further tested against the analytical $S_{\mathrm{M}}(Q)$ for homogeneous spheres provided by Botet \emph{et al.} (2020), where excellent agreement was found.
	
	\begin{figure*}
		\centering
		\includegraphics[scale=1]{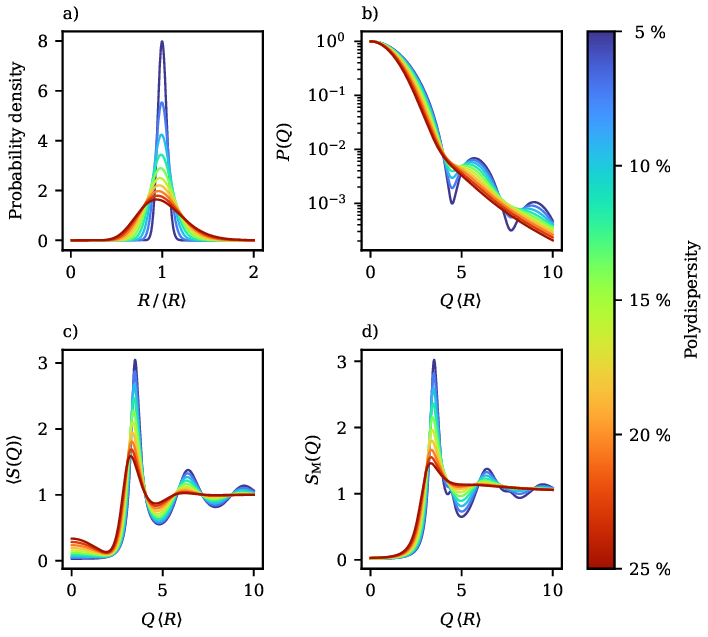}
		\caption{\label{fig:polydispersity-dependence}Comparative analysis of scattering functions for an ensemble of optically homogeneous hard spheres with varying degrees of polydispersity. Shown are: a) Probability density function illustrating the Schulz-Flory distributed-radius $R$. b) Form factor $P(Q)$. c) Average structure factor $\langle S(Q) \rangle$. d) Measurable structure factor $S_\mathrm{M}(Q)$, all evaluated at a total volume fraction of $\varphi=0.5$, spanning polydispersities from 5\,\% to 25\,\%. $\langle R \rangle$ denotes the mean radius of the spheres.}
	\end{figure*}

	\section{Measurable structure factors of polydisperse systems}
	\subsection{General remarks}
	Fig.~\ref{fig:polydispersity-dependence} provides a general overview of the influence of the polydispersity on $P(Q)$, $S_\mathrm{M}(Q)$ and $\langle S(Q) \rangle$, exemplarily discussed for a dense suspension of homogeneous spheres. Concerning the form factors, only those corresponding to polydispersities less than 10\,\% appear structured. Familiarly, the characteristic minima in $P(Q)$ become increasingly smeared out for broader size distributions. 
	
	Polydispersity also causes a change of the initial slope of $P(Q)$ in the Guinier region. Reflecting the distribution of particle sizes when calculating the Taylor expansion of $P(Q)$, the slope is now given by $-Q^2 \langle R_\mathrm G ^2 \rangle / 3$, where the familiar radius of gyration $R_\mathrm G$ is substituted by an apparent radius of gyration $\langle R_\mathrm G ^2 \rangle^{1/2}$ (Glatter, 2018; Tomchuk \emph{et al.}, 2014). For homogeneous spheres, 
	\begin{align}
		\langle R_\mathrm G ^2 \rangle = \dfrac{3}{5} \dfrac{\langle R^8 \rangle}{\langle R^6 \rangle}    
	\end{align}
	is obtained, which reduces to the well-known result of $R_\mathrm G ^2 = (3/5) R^2$ for monodisperse systems.
	
	Similar to $P(Q)$, both the measurable structure factor $S_\mathrm{M}(Q)$ and the average structure factor $\langle S(Q) \rangle$ become increasingly featureless at high polydispersities, distinctively noticeable as the principal peak's amplitude decreases and the secondary oscillations gradually disappear. Shifting the focus to the direct comparison between the two structure factors $S_\mathrm{M}(Q)$ and $\langle S(Q) \rangle$, multiple observations are apparent: While the amplitude of the principal peak is similar for both functions, differences appear at larger wavevectors, where secondary peaks in $S_\mathrm{M}(Q)$ appear at roughly the locations of the form-factor-minima, similarly noticed by Ginoza and Yasutomi (1999). With increasing polydispersity, these maxima evolve into broad shoulders which get smeared out eventually. As also noted by Ginoza and Yasutomi (1999), sharp secondary maxima are hard to observe experimentally because a very narrow size distribution in combination with a homogenous distribution of the scattering length density (SLD) inside the particles is required. Shoulder-like features in experimentally determined structure factors are on the other hand well documented, see as an example (Di Cola \emph{et al.}, 2009). In the low-$Q$ region, a striking observation is the significant increase of $\langle S(0) \rangle$ at elevated polydispersities in comparison to $S_\mathrm{M}(Q)$. According to the fluctuation-dissipation theorem from statistical mechanics, the isothermal compressibility $\kappa_\mathrm T$ is for monodisperse systems connected to the zero-wavevector limit of $S(Q)$ via $S(0) = \rho k_\mathrm B T \kappa_\mathrm T$, where $\rho$ denotes the number density and $k_\mathrm B T$ the thermal energy. The extension of this concept to mixtures must however be treated with caution, because for multi-component systems, the connection between structure and thermodynamics is not simply given by the size average $\langle S(0) \rangle$, but according to the Kirkwood-Buff theory of solutions instead via the relation $(\rho k_\mathrm B T \kappa_\mathrm T)^{-1} = \sum x_\alpha x_\beta S^{-1}_{\alpha\beta}(0)$, where $S^{-1}_{\alpha\beta}(Q)$ is the $\alpha\beta$-element of the inverse structure factor matrix (Hansen \& McDonald, 2013).
	
	\subsection{Linear contrast gradient}
	
	\begin{figure*}
		\centering
		\includegraphics[scale=1]{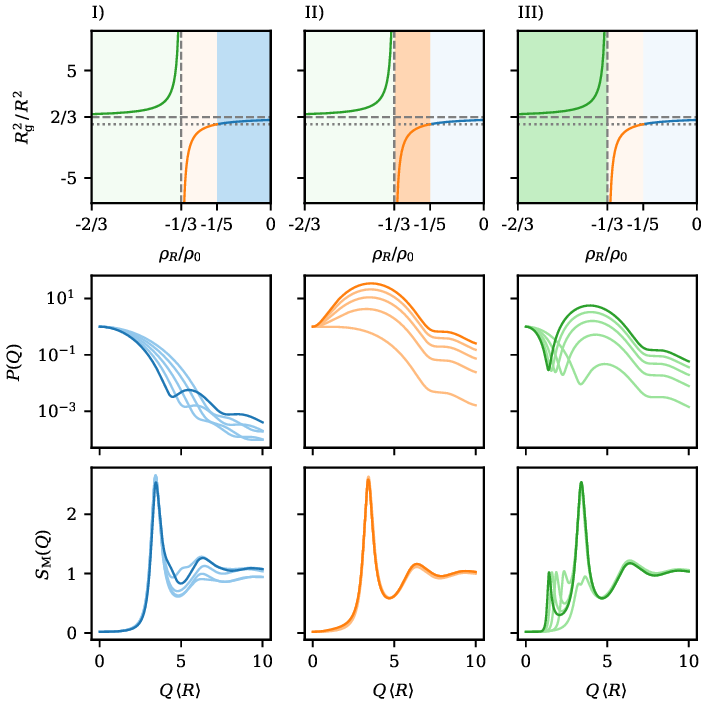}
		\caption{\label{fig:gradient-overview}Illustrative breakdown of the classification of the scattering functions of spheres with a linear contrast gradient into the three regimes discussed in the text, where each column corresponds to a unique region. In the top row, the reduced, squared radius of gyration $R_\mathrm G^2/R^2$ in dependence on the contrast ratio $\rho_R / \rho_0$ is depicted. The location of the respective region labelled I), II) or III) is indicated by the darker shaded area. The middle and bottom rows display selected form factors $P(Q)$ and measurable structure factors $S_\mathrm M(Q)$ that exemplify each region's variability in shapes observed during contrast variation. $\langle R \rangle$ indicates the mean radius of the particles. Note the shared axes of $P(Q)$ and $S_\mathrm M(Q)$ between rows and columns.}
	\end{figure*}
	
	As a prototypical example for particles with inhomogeneous scattering strength, particles with a linear gradient of the SLD are investigated. This is particularly relevant for swellable particles into which the suspension medium can diffuse. This can occur with microgel particles (Karg \emph{et al.}, 2019), as an example. Under certain reaction conditions, an inhomogeneous degree of cross-linking arises, which also leads to inhomogeneous scattering properties. Particles with intrinsic material gradients are also plausible, for example by continuously changing the monomer composition in a feed process during synthesis. Then, in principle, a suspension in which the contrast within a particle changes its sign can also be realised. The form of a linear gradient is assumed for the sake of simplicity in order to investigate the phenomenology of continuous contrasts as an example.
	
	The scattering contrast in dependence of the distance $r$ from the centre can for a single particle be parametrised as
	\begin{align}\label{eqn:gradient-contrast-profile}
		\rho(r) &=
		\begin{cases}
			\rho_{0} + (\rho_R - \rho_{0}) \dfrac{r}{R}, &\text{if $0 \leq r \leq R$},\\
			0, & \text{otherwise},\\
		\end{cases}
	\end{align}
	where $R$ is the particle radius, $\rho_0$ is the contrast in the centre and $\rho_R$ is the contrast at the interface to the surrounding medium. Accordingly, the resulting single-particle scattering amplitude is given by
	\begin{align}\label{eqn:gradient-amplitude}
		f(Q) = 4\pi \left[ \rho_0 \frac{\sin(Q R) - Q R \cos(Q R)}{Q^3} + \frac{\rho_R-\rho_0}{R} \right. \notag\\
		\left. \times \frac{2 Q R \sin(Q R) - [(Q R)^2-2] \cos(Q R) - 2}{Q^4} \right],
	\end{align}
	which reduces to 
	\begin{align}\label{eqn:gradient-forward-scattering}
		f(0) = \pi R^3 \left(\frac{\rho_0}{3} + \rho_R\right)
	\end{align}
	in the forward scattering limit. A closer look at Eq.~(\ref{eqn:gradient-forward-scattering}) reveals that the forward scattering contribution disappears if the condition $\rho_R/\rho_0 = -1/3$ is fulfilled. Especially when the maximum accessible scattering vector is limited, as in the case of light scattering, forward scattering contributes significantly to the total scattering cross-section. If the forward scattering is zero, a sample appears almost optically transparent. Refractive index matching can be achieved for particles with a homogeneous scattering capacity if the SLD of the suspension medium is adapted to that of the particles. If the scattering capacity is inhomogeneous, index matching can only minimise the total scattering cross-section, which is often achieved by making the forward scattering almost zero. In the following, the condition when the forward scattering power is minimal is referred to as the index-match-point.
	
	\begin{figure*}
		\centering
		\includegraphics[scale=1]{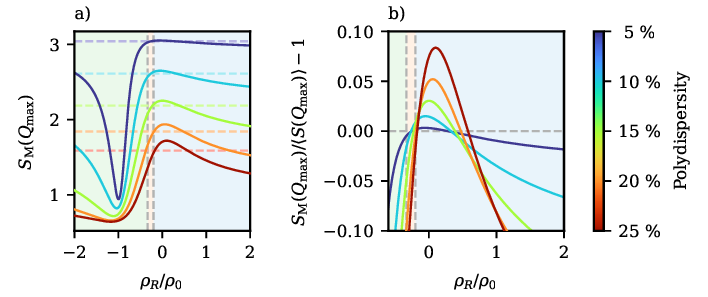}
		\caption{\label{fig:structure-factor-maximum}a) Influence of the contrast ratio $\rho_R/\rho_0$ on the principal peak value $S_\mathrm M(Q_\mathrm{max})$ for spheres with a linear contrast gradient, for polydispersities in a range between 5\,\% and 25\,\% at a total volume fraction of $\varphi=0.5$. The horizontal dashed lines mark for comparison the height of the principal peak of the average structure factor $\langle S(Q_\mathrm{max}) \rangle$. The distinction between the different introduced contrast regimes from Fig.~\ref{fig:gradient-overview} is indicated by the vertical dashed lines. b) Relative deviation between $S_\mathrm M(Q_\mathrm{max})$ and $\langle S(Q_\mathrm{max}) \rangle$ for an enlarged region.}
	\end{figure*}
	
	To gain a systematic understanding of the behaviour of the measurable structure factor $S_\mathrm M(Q)$ in dependence of the contrast ratio $\rho_R/\rho_0$, it will prove advantageous to investigate the Guinier region of the form factor. Using the contrast profile from Eq.~(\ref{eqn:gradient-contrast-profile}), for a single particle with radius $R$,
	\begin{align}\label{eqn:gradient-radius-of-gyration}
		R_\mathrm G ^2 = \dfrac{2}{5} \, \dfrac{\rho_0+5 \rho_R}{\rho_0+3\rho_R} \,R^2
	\end{align}
	is obtained for the effective squared radius of gyration, which depends besides the particle's radius also on the two contrast parameters $\rho_0$ and $\rho_R$. For polydisperse suspensions, a similar expression emerges:
	\begin{align}\label{eqn:gradient-radius-of-gyration-polydisperse}
		\langle R_\mathrm G ^2 \rangle = \dfrac{2}{5} \, \dfrac{\rho_0+5 \rho_R}{\rho_0+3\rho_R} \, \dfrac{\langle R^8 \rangle}{\langle R^6 \rangle}.
	\end{align}
	As such, the contrast-dependence of the prefactor is not altered by polydispersity and the qualitative discussion can instead be based on monodisperse suspensions. We will thus refer to the prefactor simply as $R_\mathrm G ^2 / R^2$, even in the polydisperse case. 
	
	Inspecting Eq.~(\ref{eqn:gradient-radius-of-gyration}), several characteristic ratios $\rho_R/\rho_0$ are apparent: $R_\mathrm G ^2$ becomes zero for $\rho_R/\rho_0 = -1/5$, exhibits a pole at $\rho_R/\rho_0 = -1/3$, incident with the index-match-point and has an asymptotic limit of $R_\mathrm G ^2/R^2 = 2/3$ for $\rho_R/\rho_0 \to \pm\infty$. It will be shown that the behaviour of the scattering functions can be divided into three qualitatively distinct classes and that  form factors and measurable structure factors within each domain share unique features. The classification based on the behaviour of $R_\mathrm G ^2$, together with form factors $P(Q)$ and measurable structure factors $S_\mathrm M(Q)$ representative for each region is visualised in Fig.~\ref{fig:gradient-overview}. The regions are characterised as follows:
	
	(I) For $\rho_R/\rho_0 > -1/5$, $R_\mathrm G ^2$ is positive and the form factors have the familiar decaying shape known from homogeneous spheres. With decreasing contrast ratio, the decay becomes increasingly gradual until $R_\mathrm G ^2=0$ is reached for $\rho_R/\rho_0 = -1/5$. Around the principal peak of $S_\mathrm M(Q)$ and for lower wavevectors, changes in the contrast have a negligible influence on the measurable structure factors. Contrarily, at wavevectors beyond the principal peak's location, $S_\mathrm M(Q)$ is greatly affected by contrast variation. Depending on the specific location of the first form factor minimum, which shifts to larger wavevectors with lower contrast ratios, the shoulder-like artefact also visible in Fig.~\ref{fig:polydispersity-dependence} moves through $S_\mathrm M(Q)$ towards larger wavevectors and therein most prominently affects the shape of the first local minimum and the following secondary maximum.
	
	(II) For the contrast ratios $-1/3 < \rho_R/\rho_0 < -1/5$, $R_\mathrm G ^2$ becomes negative, which implies an imaginary radius of gyration $R_{\rm G}$ leading to a positive initial slope of $P(Q)$. Form factors in this region therefore initially increase from $P(0)=1$ until a global maximum is reached at $QR\approx4$, after which they decay. The height of the maximum increases as the contrast ratio moves towards the index-match-point at $\rho_R/\rho_0 = -1/3$. Curiously, the measurable structure factors in this domain are almost indistinguishable, even though the variation of $R_\mathrm G ^2$ is much more pronounced in comparison to region (I), where the span of $R_\mathrm G ^2$ is small, but $S_\mathrm M(Q)$ shows a much more diverse behaviour. Also, the distorting artefacts from region (I) disappear almost completely.
	
	(III) Contrast ratios $\rho_R/\rho_0 < -1/3$ again result in positive $R_\mathrm G ^2$ and negative initial slopes. Close to the index-match-point, where $R_\mathrm G ^2$ is comparatively large, $P(Q)$ exhibits an intriguing shape: At small wavevectors, an unusually sharp minimum occurs even in very polydisperse suspensions. Beyond the minimum, $P(Q)$ rises to a global maximum reminiscent of region (II). For contrasts in this range, an additional local maximum in $S_\mathrm M(Q)$ appears at low wavevectors, caused by the presence of the first form factor minimum. Such secondary maxima are often discussed in the literature as an indication of self-organisation on length scales beyond the distance of nearest neighbours, i.e., the formation of correlated clusters (Sciortino \emph{et al.} 2004; Liu \emph{et al.}, 2005). The secondary maxima occurring here are exclusively caused by the scattering amplitudes and cannot be attributed to structural properties of the sample. This constitutes a valuable example for a situation where a careless inspection of experimentally determined $S_\mathrm M(Q)$ can in the worst case lead to unjustified assumptions about the structure of a system. Moving further away from the index-match-point, the first form factor minimum moves towards larger wavevectors and gets shallower. At the same time, the following maximum declines and as such, the shape of $P(Q)$ morphs back into the familiar decaying shape from region (I). Simultaneously, the location of the secondary maximum in $S_\mathrm M(Q)$ drifts towards higher wavevectors. Fig.~\ref{fig:gradient-overview} also displays a situation where the form factor minimum exactly coincides with the location where the principal peak of $S_\mathrm M(Q)$ would normally occur. In this case the main peak is drastically diminished, which is again not an indicator for a less pronounced short-range order in this particular instance, but can certainly be mistaken as such.
	
	The principal peak height of a structure factor is an often employed structural order parameter. Scheffold and Mason (2009) noticed in their investigation of highly concentrated nanoemulsions that the peak amplitude in $S_\mathrm M(Q)$ is deeply affected by polydispersity. As such, also the evolution of this height during contrast variation is of special interest. Fig.~\ref{fig:structure-factor-maximum} compares the peak height of the average structure factor $\langle S(Q_\mathrm{max}) \rangle$ to the value of $S_\mathrm M(Q_\mathrm{max})$ at the same wavevector in dependence on the contrast ratio $\rho_R/\rho_0$ and for different degrees of polydispersity. Overall, it is clearly shown that $S_\mathrm M(Q_\mathrm{max})$ is deeply affected by changes in the contrast. There exist two contrast ratios where $S_\mathrm M(Q_\mathrm{max})$ and $\langle S(Q_\mathrm{max}) \rangle$ coincide. One of them is to a good approximation given by $\rho_R/\rho_0 \approx -1/5$, the location where the apparent radius of gyration disappears and $P(Q)$ decays very slowly. The other location is at a positive contrast ratio and drifts towards higher $\rho_R/\rho_0$ with increasing polydispersity. Bounded by those two ratios is a regime where $S_\mathrm M(Q_\mathrm{max})$ exceeds $\langle S(Q_\mathrm{max}) \rangle$, while for all other contrast ratios, the peak height from $S_\mathrm M(Q_\mathrm{max})$ underestimates the actual height. For comparatively small polydispersities around $5\,\%$, the deviation from $\langle S(Q_\mathrm{max}) \rangle$ is small and only amounts to a few percent, as long as the contrast ratio is larger than $\rho_R/\rho_0 \approx -1/5$. For lower ratios, $S_\mathrm M(Q_\mathrm{max})$ is strongly diminished, most pronounced at contrast ratios of $\rho_R/\rho_0 \approx -1$. For higher polydispersities, the deviations become even more severe, as best visualised in Fig.~\ref{fig:structure-factor-maximum} b), where the relative deviation between $\langle S(Q_\mathrm{max}) \rangle$ and $S_\mathrm M(Q_\mathrm{max})$ is depicted. Even in the immediate vicinity of $\langle S(Q_\mathrm{max}) \rangle = S_\mathrm M(Q_\mathrm{max})$, already deviations in the order of $5-10\,\%$ appear for the highest shown polydispersities. This shows that --- no matter the actual degree of polydispersity --- $S_\mathrm M(Q_\mathrm{max})$ can only serve as a reliable order parameter for very specific contrast ratios.
	
	\subsection{Core-shell particles}
	Core-shell models are commonly employed to describe particles consisting of different layers of material, for example nanoparticles with grafted stabiliser shells (Hallet \emph{et al.}, 2020; Diaz Maier \& Wagner, 2024) or micellar structures (Szymusiak, 2017). As core and shell naturally differ in their material properties, in principle both positive and negative contrast differences with respect to the surrounding medium can occur, similar to particles with continuous material gradients. For Schulz-Flory distributed core-shell particles, analytical expressions for the form factor $P(Q)$ exist in the case of a polydisperse core and a shell of constant thickness (Bartlett \& Ottewill, 1992), a polydisperse total diameter and a constant core-to-shell ratio (Wagner, 2004) and for both core radius and shell thickness independently distributed  (Wagner, 2012). Moreover, an analytical solution for the problem of correlated hard-sphere core-shell systems was provided by Nayeri \emph{et al.} (2009).
	
	The scattering amplitude of a single core-shell particle 
	\begin{align}\label{eqn:core-shell-amplitude}
		f(Q) = 4\pi \left[ (\rho_\mathrm c - \rho_\mathrm s) \dfrac{\sin(QR_{\mathrm c}) -QR_{\mathrm c} \cos(QR_{\mathrm c})}{Q^3} \right.\notag \\
		+ \left. \rho_\mathrm s \dfrac{\sin(QR) - QR \cos(QR)}{Q^3} \right]
	\end{align}
	is the sum of the amplitudes of a sphere and a spherical shell, weighted by their respective contrasts, $\rho_\mathrm c$ and $\rho_\mathrm s$. $R_{\mathrm c}$ and $R$ are, respectively, the core radius and the total radius of the particle and we specifically consider the case where the core radius and the total radius are connected by a constant, species-independent size ratio $\delta = R_{\mathrm c}/R$.
	
	Similar to the gradient model, the forward scattering contribution
	\begin{align}\label{eqn:core-shell-forward-amplitude}
		f(0) = \dfrac{4}{3} \pi R^3 \left[ \delta^3 (\rho_\mathrm c - \rho_\mathrm s) + \rho_\mathrm s \right]
	\end{align}
	disappears for specific contrast combinations $\rho_\mathrm s/\rho_\mathrm c = \delta^3 / (\delta^3 - 1)$, where the ratio of contrasts now additionally depends on the size ratio $\delta$. For the effective radius of gyration of a polydisperse system, an expression with similar structure to Eq.~(\ref{eqn:gradient-radius-of-gyration-polydisperse}) emerges:
	\begin{align}\label{eqn:core-shell-radius-of-gyration-polydisperse}
		\langle R_\mathrm G ^2 \rangle = \dfrac{3}{5} \, \dfrac{\delta^5 \rho_\mathrm c + (1-\delta^5) \rho_\mathrm s}{\delta^3 \rho_\mathrm c + (1-\delta^3) \rho_\mathrm s} \,\dfrac{\langle R^8 \rangle}{\langle R^6 \rangle}.
	\end{align}
	That again, a prefactor containing the contrasts can be decoupled from the size average is a peculiarity of this model with constant size ratio and a key reason why this assumption was made for this investigation.
	
	\begin{figure}
		\centering
		\includegraphics[width=\linewidth]{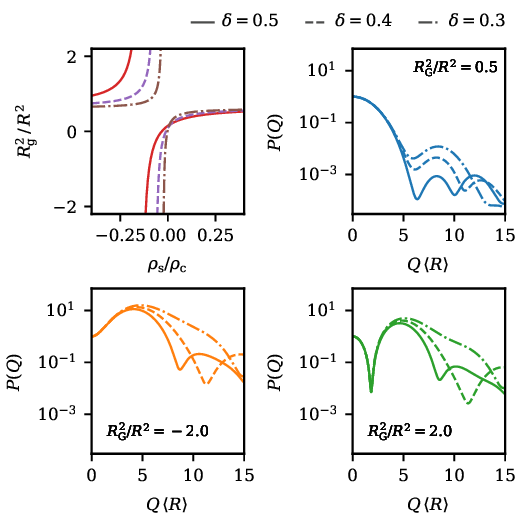}
		\caption{\label{fig:core-shell-contrast}Influence of the contrast ratio $\rho_\mathrm s / \rho_\mathrm c$ on the reduced squared radius of gyration $R_\mathrm G ^2 / R^2$ for core-shell particles with different ratios $\delta$ between core radius and total radius, along with three representative sets of form factors, each sharing the same radius of gyration for different size ratios.}
	\end{figure}
	In Fig.~\ref{fig:core-shell-contrast}, the contrast-dependence of $R_\mathrm G ^2 / R^2$ is visualised for different size ratios $\delta$. As in the case of spheres with a linear gradient of the SLD, this results in hyperbola-like curves, where the location of the pole is now influenced by $\delta$: An increasing ratio of core diameter to total diameter shifts the location of the pole to more negative contrast ratios $\rho_\mathrm s / \rho_\mathrm c$. The contrast ratio where $R_\mathrm G ^2 = 0$ is in comparison only slightly altered by $\delta$. This leads to a larger range of contrast ratios with negative $R_\mathrm G ^2$ as the shell thickness decreases. 
	
	This shows that core-shell particles exhibit qualitatively comparable optical characteristics to particles with a linear density gradient. As such, the form factors $P(Q)$ of core-shell systems can likewise be categorised into three classes based on their behaviour at low wavevectors. Exemplary form factors for each class are also visualised in Fig.~\ref{fig:core-shell-contrast}.
	
\begin{figure}[h]
		\centering
		\includegraphics[width=\linewidth]{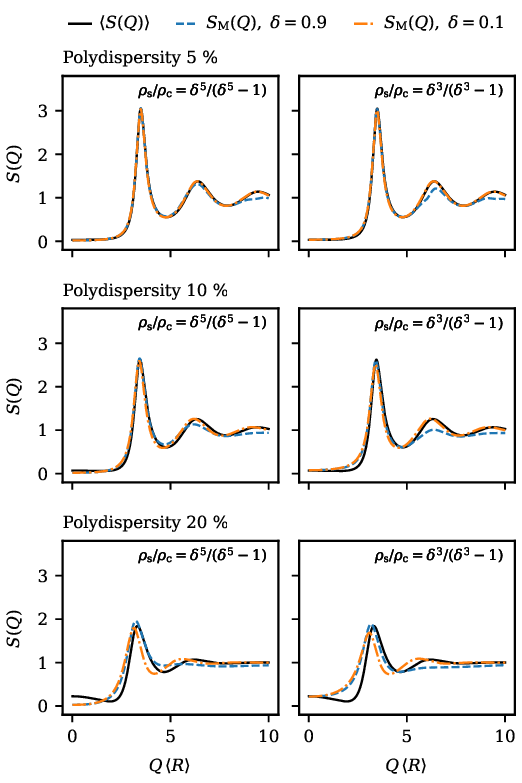}
		\caption{\label{fig:core-shell-structure-factor}Comparison between measurable structure factor $S_\mathrm M(Q)$ and size-averaged structure factor $\langle S(Q) \rangle$ of core-shell particles for different core-to-total ratios $\delta$, contrast ratios $\rho_\mathrm s/\rho_\mathrm c$ and polydispersities as indicated in the figure. The total volume fraction for all shown structure factors is $\varphi=0.5$.}
	\end{figure}

	Because of these similarities, we focus the remainder of the discussion on aspects that are unique to core-shell particles, i.e., how measurable structure factors are influenced by different core-to-shell ratios. For this purpose, exemplarily, structure factors corresponding to two important edge-cases, particles with a small core and particles with a thin shell, are compared in Fig.~\ref{fig:core-shell-structure-factor} for different degrees of polydispersity and chosen contrast ratios $\rho_\mathrm s/\rho_\mathrm c$. Core-shell models with thin shells are often encountered when characterising particles stabilised by a grafted polymer layer, which are prototypical colloidal model particles displaying hard-sphere behaviour (Royall \emph{et al.}, 2013). The case of hard spheres with a strongly scattering, small core, and a weakly scattering, comparatively large shell is equally of interest: Under these conditions, essentially the behaviour of highly charged, strongly repelling particles whose interparticle distance is several times larger than their diameter, is artificially mimicked. For these systems, the measurable structure factor $S_\mathrm M(Q)$ should in theory to a good approximation coincide with the average structure factor $\langle S(Q) \rangle$. To reasonably compare models with different size ratios $\delta$, two specific contrast ratios $\rho_\mathrm s/\rho_\mathrm c$ are depicted, the ratio $\rho_\mathrm s/\rho_\mathrm c = \delta^3 / (\delta^3-1)$ at the index matching point, where forward scattering is minimised, and the ratio $\rho_\mathrm s/\rho_\mathrm c = \delta^5 / (\delta^5-1)$, where $\langle R_\mathrm G ^2 \rangle = 0$ and $P(Q)$ shows the weakest decay. In the case of $\delta \ll 1 $, both conditions basically lead to the same result: The shell is virtually hidden with $\rho_\mathrm s \approx 0$. 
	
	As can be observed in Fig.~\ref{fig:core-shell-structure-factor}, for moderate polydispersities of 5 - 10 \%, the small core-to-total ratio $\delta=0.1$ indeed yields measurable structure factors $S_\mathrm M(Q)$ which are indistinguishable from $\langle S(Q) \rangle$ for both depicted contrast ratios. For particles with thin shells ($\delta=0.9$), $S_\mathrm M(Q)$ and $\langle S(Q) \rangle$ also agree well in the vicinity of the principal peak. However, differences arise around the secondary maxima, where the peak amplitudes in $S_\mathrm M(Q)$ are diminished as a cause of the interference of the scattering amplitudes. With increasing polydispersity, this deviation becomes more pronounced. Still, even for particles which are seemingly quite close to homogeneous spheres, artefacts in $S_\mathrm M(Q)$ can be significantly reduced by careful contrast variation. Looking at highly polydisperse systems, it is evident that even for rather small cores with $\delta=0.1$, $\langle S(Q) \rangle$ cannot be accurately represented by any $S_\mathrm M(Q)$. Only the height of the principle peak is correctly estimated. This stresses again the importance of an accurate treatment of very broad size distributions, where any kind of approximation must be carefully checked for validity.
	
	\section{Conclusions}
	Colloidal dispersions generally exhibit a particle size distribution which needs to be taken into account when interpreting results from scattering experiments. The measurable structure factor $S_\mathrm M(Q)$ is an experimental, comparatively easy accessible measure for the interparticle structure in interacting systems. However, it is important to be aware that in polydisperse systems, $S_\mathrm M(Q)$ is beyond the structural correlations also decisively affected by the optical properties of the individual particles. To this end, we systematically investigate the influence of different form factor models on the shape of $S_\mathrm M(Q)$ of dense dispersions with hard-sphere interactions. The characterisation of measurable structure factors  is extended to two classes of spherical particles with inhomogeneous scattering capacity: First, spheres with a linear SLD-profile as a general model for particles with continuous contrast gradients and second, a core-shell-system as a prototype for particles with layered structures. 
	
	For both models, we find that the structure factors can be categorised into three distinctive classes of shared qualitative features, based on the behaviour of the form factor $P(Q)$ in the Guinier region. $S_\mathrm M(Q)$ can for these optically inhomogeneous model particles be significantly influenced by the variation of the scattering contrasts relative to the surrounding medium. Depending on the specific contrast combination, shoulder-like features emerge, maxima are diminished or split and even secondary maxima in the low wavevector region, reminiscent of cluster-peaks, can be observed. These effects are solely due to the optical properties of the particles and are not caused by structural changes in the sample. We further show that the height of the principal peak of $S_\mathrm M(Q)$ can only be regarded as a representative order parameter in a very restricted range of contrasts, especially for broad size distributions.
	
	These observations emphasise the need to properly address the distribution of particle size (and possibly also other characteristics) in the interpretation of static scattering experiments. Actually, for many applications, deliberately broad size distributions are a desired feature, an academically relevant example being studies of deeply supercooled, glass-forming systems (Ninarello \emph{et al.}, 2017), where crystallisation needs to be suppressed and where polydispersity effects in any form certainly cannot be neglected (Zaccarelli \emph{et al.}, 2015; Pihlajamaa \emph{et al.}, 2023).
	
	Beyond providing an enhanced qualitative understanding of features which can possibly be encountered when analysing experimentally extracted measurable structure factors, the numerical scheme presented in this contribution in principle provides a means to model the scattered intensity of any polydisperse hard-sphere system, provided a model for the single-particle scattering amplitude and an appropriate size distribution is available. Performing fits with such advanced models directly on experimentally observed intensities gives access to the underlying partial structure factors, enabling a characterisation and possible further theoretical analysis on a genuine multi-component foundation, rather than employing effective one-component approaches. The current restriction to hard-sphere interactions is a major incentive to promote advancements in the analytical evaluation of partial structure factors for other interaction potentials, since numerically solving integral equations or employing computer simulations with reasonable statistics is currently only realistically feasible for a restricted number of components, especially in mixtures with large size disparities (Allahyarov \emph{et al.}, 2022).
	
	\section*{References}
	\reference{Allahyarov, E., Löwen, H. \& Denton, A. R. (2022). \emph{Phys. Chem. Chem. Phys.} \textbf{24}, 15439--15451.}
	\reference{Ballauff, M. (2001). \emph{Curr. Opin. Colloid Interface Sci.} \textbf{6}, 132--139.}
	\reference{Banchio, A. J., Nägele, G., \& Ferrante, A. (1998). \emph{J. Colloid Interface Sci.}, \textbf{208}, 487-499.}
	\reference{Bartlett, P. \& Ottewill, R. H. (1992). \emph{J. Chem. Phys.} \textbf{96}, 3306--3318.}
	\reference{Baxter, R. J. (1970). \emph{J. Chem. Phys.} \textbf{52}, 4559--4562.}
	\reference{Blum, L. \& Stell, G. (1979). \emph{J. Chem. Phys.} \textbf{71}, 42--46.}
	\reference{Blum, L. \& Stell, G. (1980). \emph{J. Chem. Phys.} \textbf{72}, 2212.}
	\reference{Bohren, C. F. \& Huffman, D. R. (2008). \emph{Absorption and Scattering of Light by Small Particles}. New York: John Wiley.}
	\reference{Botet, R., Kwok, S. \& Cabane, B. (2020). \emph{J. Appl. Crystallogr.} \textbf{53}, 1570--1582.}
	\reference{D'Aguanno, B. \& Klein, R. (1992). \emph{Phys. Rev. A}, \textbf{46}, 7652--7656.}
	\reference{D'Aguanno, B. (1993). \emph{Phys. Scr.} \textbf{T49}, 84--88.}
	\reference{Di Cola, E., Moussaïd, A., Sztucki, M., Narayanan, T. \& Zaccarelli, E. (2009). \emph{J. Chem. Phys.} \textbf{131}, 144903-1--144903-9.}
	\reference{Diaz Maier, J. \& Wagner, J. (2024). \emph{Soft Matter}, \textbf{20}, 1309--1319}
	\reference{Flory, P. J. (1936). \emph{J. Am. Chem. Soc.} \textbf{58}, 1877--1885.}
	\reference{Ginoza, M. \& Yasutomi, M. (1999). \emph{J. Phys. Soc. Jpn.} \textbf{68}, 2292--2297.}
	\reference{Glatter, O. (2018). \emph{Scattering Methods and their Application in Colloid and Interface Science}. Amsterdam: Elsevier.}
	\reference{Griffith, W. L., Triolo, R. \& Compere, A. L. (1987). \emph{Phys. Rev. A}, \textbf{35}, 2200--2206.}
	\reference{Hansen, J.-P., Levesque, D. \& Zinn-Justin, J. (1991). \emph{Liquids, Freezing and Glass
			Transition}. Amsterdam: North Holland.}
	\reference{Hansen, J.-P. \& McDonald, I. R. (2013). \emph{Theory of Simple Liquids with Applications to Soft Matter}. Amsterdam: Elsevier.}
	\reference{Hayter, J. B. \& Penfold, J. (1981), \emph{Mol. Phys.}, \textbf{42}, 109--118}
	\reference{Hallett, J. E., Grillo, I., Smith, G. N. (2020). \emph{Langmuir}, \textbf{36}, 2071--2081.}
	\reference{Karg, M., Pich, A., Hellweg, T., Hoare, T., Lyon, L. A., Crassous, J. J., Suzuki, D., Gumerov, R. A., Schneider, S., Potemkin, I. I., Richtering, W. (2019). \emph{Langmuir}, \textbf{35}, 6231--6255.}
	\reference{Kirkwood, J. G., Boggs, E. M. (2004). \emph{J. Chem. Phys.}, \textbf{10}, 394--402.}
	\reference{Kotlarchyk, M., Chen, S.-H. (1983). \emph{J. Chem. Phys.}, \textbf{79}, 2461--2469.}
	\reference{Li, T., Senesi, A. J., \& Lee, B. (2016). \emph{Chem. Rev.}, \textbf{116}, 11128--11180.}
	\reference{Liu, Y., Chen, W.-R., Chen, S.-H. (2005). \emph{J. Chem. Phys.}, \textbf{122}, 044507.}
	\reference{Lu, P. J., \& Weitz, D. A. (2013). \emph{Annu. Rev. Condens. Matter Phys.} \textbf{4}, 217--233.}
	\reference{Nayeri, M., Zackrisson, M., \& Bergenholtz, J. (2009). \emph{J. Phys. Chem. B}, \textbf{113}, 8296--8302}
	\reference{Ninarello, A., Berthier, L. \& Coslovich, D. (2017). \emph{Phys. Rev. X}, \textbf{7}, 021039}
	\reference{Olver, F. W. J., Lozier, D. W., Boisvert, R. F., \& Clark, C. W. (2010). \emph{The NIST Handbook of Mathematical Functions}. Cambridge: Cambridge University Press.}
	\reference{Percus, J. K. \& Yevick, G. J. (1958). \emph{Phys. Rev.} \textbf{110}, 1--13.}
	\reference{Pihlajamaa, I., Laudicina, C. C. L. \& Janssen, L. M. C. (2023). \emph{Phys. Rev. Res.} \textbf{5}, 033120}
	\reference{Pusey, P. N., Fijnaut, H. M., \& Vrij, A. (1982). \emph{J. Chem. Phys.}, \textbf{77}, 4270--4281.}
	\reference{Royall, C. P., Poon, W. C. K., Weeks, E. R. (2013). \emph{Soft Matter}, \textbf{9}, 17--27.}
	\reference{Salgi, P. \& Rajagopalan, R. (1993). \emph{Adv. Colloid Interface Sci.} \textbf{43}, 169--288.}
	\reference{Scheffold, F. \& Mason, T. G. (2009). \emph{J. Phys.: Condens. Matter} \textbf{21}, 332102.}
	\reference{Schulz, G. V. (1939). \emph{Z. Phys. Chem.} \textbf{43b}, 25--46.}
	\reference{Sciortino, F., Mossa, S., Zaccarelli, E., Tartaglia, P. (2004). \emph{Phys. Rev. Lett.}, \textbf{93}, 055701.}
	\reference{Szymusiak, M., Kalkowski, J., Luo, H., Donovan, A. J., Zhang, P., Liu, C., Shang, W., Irving, T., Herrera-Alonso, M., Liu, Y. (2017). \emph{ACS Macro Lett.}, \textbf{6}, 1005--1012.}
	\reference{Tomchuk, O. V., Bulavin, L. A., Aksenov, V. L., Garamus, V. M., Ivankov, O. I., Vul', A. Y., Dideikin. A. T. \& Avdeev, M. V. (2014). \emph{J. Appl. Cryst.}, \textbf{47}, 643--653.}
	\reference{van Beurten, P. \& Vrij, A. (1981). \emph{J. Chem. Phys.} \textbf{74}, 2744--2748.}
	\reference{Voigtmann, T. (2003). \emph{Mode Coupling Theory of the Glass Transition in Binary Mixtures}, Technische Universität München}
	\reference{Vrij, A. (1978). \emph{J. Chem. Phys.} \textbf{69}, 1742--1747.}
	\reference{Vrij, A. (1979). \emph{J. Chem. Phys.} \textbf{71}, 3267--3270.}
	\reference{Wagner, J. (2004). \emph{J. Appl. Cryst.} \textbf{37},  750--756.}
	\reference{Wagner, J. (2012). \emph{J. Appl. Cryst.} \textbf{45},  513--516.}
	\reference{Widom, B. (1967). \emph{Science}, \textbf{157} (3787), 375--382.}
	\reference{Zaccarelli, E., Liddle, S. M. \&  Poon, W. C. K. (2015). \emph{Soft Matter}, \textbf{11}, 324--330.}
	\reference{Zhou, B., Gasser, U. \& Fernandez-Nieves, A. (2023). \emph{Phys. Rev. E}, \textbf{108}, 054604.}
	
	\appendix
	\section{Percus-Yevick structure factors for hard sphere mixtures}
	The analytical solution of the Ornstein-Zernike equation for the hard-sphere potential within the Percus-Yevick closure in terms of the partial structure factors $S_{\alpha\beta}(Q)$, presented by Vrij (1979) and reformulated by Voigtmann (2003), is restated here. In short, an expression for the partial direct correlation functions $c_{\alpha\beta}(r)$ in real space can be found using Baxter's (1970) factorisation technique. The transformed solution in wavevector-space $c_{\alpha\beta}(Q)$ can subsequently be used to obtain the partial structure factors $S_{\alpha\beta}(Q)$.
	
	Let $\varphi$ be the total volume fraction of all spheres, $d_\alpha$ be the diameter and $x_\alpha$ be the number fraction of the spheres of species $\alpha$. The total number density $\rho$ of the system is related to the volume fraction by $\varphi=(\pi/6) \rho \sum x_\alpha d_\alpha^3$. 
	With the abbreviations
	\begin{align}
		d_{\alpha\beta} &= \dfrac{d_\alpha + d_\beta}{2},\\
		\overline{d}_{\alpha\beta} &= \dfrac{d_\alpha - d_\beta}{2},\\
		\xi_x &= \dfrac{\pi}{6} \rho \sum_\gamma x_\gamma d_\gamma^x,
	\end{align}
	
	the set of coefficients
	\begin{align}
		a_\alpha &= \frac{1 - \xi_3 + 3d_\alpha \xi_2}{\left(1-\xi_3 \right)^2}\\
		b_\alpha &= - \frac{3}{2} \frac{d_\alpha^2 \xi_2}{(1-\xi_3)^2}\\
		\Tilde{a}_2 &= \sum \limits_\gamma \rho_\gamma a_\gamma^2 \\
		\hat{\beta_0} &= \frac{9\xi_2^2 + 3\xi_1(1-\xi_3)}{(1-\xi_3)^3}\\
		A_{\alpha\beta} &= \frac{d_{\alpha \beta} ( 1 -\xi_3) + \frac{3}{2}d_\alpha d_\beta \xi_2}{(1-\xi_3)^2} \\
		B_{\alpha \beta} &= \frac{1}{1-\xi_3} - \hat{\beta}_0 d_\alpha d_\beta \\
		D_{\alpha \beta } &= \frac{6\xi_2 + 12d_{\alpha \beta}(\xi_1 + 3\xi_2^2 / (1-\xi_3))}{(1-\xi_3)^2}
	\end{align}
	can be determined. Introducing further $S_\alpha = \sin(Q d_\alpha / 2)$ and $C_\alpha = \cos(Q d_\alpha / 2)$, the terms
	\begin{align}
		\mu_A &= A_{\alpha \beta}\frac{S_\alpha S_\beta - C_\alpha C_\beta}{Q^2}\\
		\mu_B &= B_{\alpha \beta} \frac{C_\alpha S_\beta + C_\beta S_\alpha}{Q^3}\\
		\mu_D &= D_{\alpha \beta}\frac{S_\alpha S_\beta}{Q^4}
	\end{align}
	and
	\begin{align}
		\tilde\mu = \frac{4\pi}{Q^4}\tilde{a}_2 \biggl(&\frac{C_\alpha C_\beta d_\alpha d_\beta}{4} + \frac{S_\alpha S_\beta}{Q^2} \notag\\
		&- \frac{C_\alpha S_\beta d_\alpha + C_\beta S_\alpha d_\beta}{2Q} \biggr).
	\end{align}
	can be calculated, which finally leads to
	\begin{align}
		c_{\alpha \beta}(q) = -4\pi \, (\mu_A + \mu_B + \mu_D + \tilde\mu).
	\end{align}
	The partial direct correlation functions form the matrix $\mathbf{C}$ with elements $C_{\alpha\beta} = (x_\alpha x_\beta)^{1/2} c_{\alpha\beta}$, which is related to the matrix of partial structure factors $\mathbf{S}$ by the Ornstein-Zernike relation 
	\begin{align}
		\mathbf{S} = \left[ \mathbf{1} - \rho \, \mathbf{C} \right]^{-1}.
	\end{align}
	The partial structure factors here are defined within the convention $\lim\limits_{Q\to\infty} S_{\alpha\beta}(Q) = \delta_{\alpha\beta}$, where $\delta_{\alpha\beta}$ is the Kronecker delta.
	\clearpage
\end{document}